\documentclass[article,twocolumn,preprintnumbers,bibnotes10pt,superscriptaddress]{revtex4}

\usepackage{amsmath}
\usepackage{graphicx}
\usepackage{ulem}
\usepackage{dcolumn}
\usepackage{epsfig}
\usepackage{bm}
\usepackage{array}
\usepackage[frenchb]{babel}
\usepackage{hyperref}
\hypersetup{
colorlinks=true,
citecolor=black,
linkcolor=red,
urlcolor=black
 , bookmarks=true,
 pdfmenubar=true
}
\usepackage{hyperref}
\usepackage{graphicx}
\usepackage{amsmath}
\usepackage{amsfonts}
\usepackage{amssymb}
\usepackage{xcolor}
\usepackage{bbm}
\usepackage{calrsfs}
\usepackage{dutchcal}
\usepackage{amsthm}
\usepackage{xcolor}
\usepackage{wrapfig}

\usepackage[utf8]{inputenc}
\usepackage[T1]{fontenc}
\usepackage{bm}
\usepackage{physics}
\usepackage{amsfonts}
\usepackage[nonumberlist, acronym, toc]{glossaries}
\usepackage{ulem} 
\usepackage{tcolorbox}
\usepackage{wrapfig}

\newtcolorbox[use counter=myexample,number format=\Alph 
]{mybox}[2][]{%
colback=black!5!white,colframe=black!80!white,fonttitle=\bfseries,
title=Box \thetcbcounter: #2,#1}

\newcommand{\Eins}{\ensuremath{\mathbbm 1}}

\newcommand{\vect}[1]{\bm{#1}}

\newcommand{\be}{\begin{equation}}
\newcommand{\ee}{\end{equation}}

\newcommand{\beq}{\begin{eqnarray}}
\newcommand{\eeq}{\end{eqnarray}}

\newcommand{\rhotheta}{\rhoop_{\boldsymbol{\theta}}}

\newcommand{\thest}{\vect{\Theta}}

\newcommand{\rhoop}{\rho}
\newcommand{\Hop}{H}
\newcommand{\POVMop}{E}
\newcommand{\Lop}{L_{\vect{\theta}}}

\begin{document}

\title{Advances in multiparameter quantum sensing and metrology}

\author{Luca Pezz\`e}

\affiliation{Istituto Nazionale di Ottica, Consiglio Nazionale delle Ricerche (CNR-INO), Largo Enrico Fermi 6, 50125 Firenze, Italy} 
\affiliation{European Laboratory for Nonlinear Spectroscopy (LENS), Via N. Carrara 1, 50019 Sesto Fiorentino, Italy}
\affiliation{QSTAR, Largo Enrico Fermi 2, 50125 Firenze, Italy}

\author{Augusto Smerzi}

\affiliation{Istituto Nazionale di Ottica, Consiglio Nazionale delle Ricerche (CNR-INO), Largo Enrico Fermi 6, 50125 Firenze, Italy} 
\affiliation{European Laboratory for Nonlinear Spectroscopy (LENS), Via N. Carrara 1, 50019 Sesto Fiorentino, Italy}
\affiliation{QSTAR, Largo Enrico Fermi 2, 50125 Firenze, Italy}

\begin{abstract}
Recent years have witnessed a growing interest in understating the limitations imposed by quantum noise in precision measurements and devising techniques to reduce it.
The attention is currently turning to the simultaneously estimation of several parameters of interest, driven by its promising potential across a wide range of sensing applications as well as fueled by experimental progress in various optical and atomic platforms.
Here, we provide a comprehensive overview of key research directions in multiparameter quantum sensing and metrology, highlighting both opportunities and challenges. 
We introduce the basic framework, discuss ultimate sensitivity bounds, optimal measurement strategies, and the role of quantum incompatibility, showing important differences with respect to single-parameter estimation.
Additionally, we discuss emerging experimental implementations in distributed quantum sensing, including cutting-edge optimization techniques. 
This review aims to bridge the gap between theory and experiments, paving the way for the next-generation of quantum sensors and their integration with other quantum technologies.
\end{abstract}

\maketitle
\date{\today}


{\bf Introduction.} 
Multiparameter sensing focuses on the simultaneous estimation of several parameters.
This task has a broad range of applications, where the achievable precision is -- or will soon be -- fundamentally limited by the intrinsic quantum noise.
For example, quantum networks of atomic clocks~\cite{ZollerPRL2024, KomarNATPHYS2014} enhance precision measurements over large distances, enabling unprecedented synchronization and stability.
Optical imaging~\cite{AlbarelliPLA2020} and magnetic field mapping~\cite{DegenRMP2017} impact medicine and biology~\cite{BowenPR2016, AslamNRP2023}, improving monitoring and diagnostics. 
Vectorial force sensing enables high-precision inertial navigation~\cite{GracePRAPP2020}.
Learning quantum systems~\cite{GebhartNRP2023},
designing quantum gates and algorithms~\cite{BhartiRMP2022} and communication networks~\cite{WehnerSCIENCE2018} can involve the estimation and the optimization of multiple parameters.
Furthermore, the real-time simultaneous estimation of phase and phase diffusion helps reduce systematic errors~\cite{VidrighinNATCOMM2014}.


The aim of this review is to explore possibilities and challenges in multiparameter quantum sensing and metrology.
Recent theoretical and experimental studies have highlighted how quantum resources -- such as entanglement and squeezing -- can reduce the impact of quantum noise and thus enhance the performance of multiparameter estimation beyond what is possible using independent sensors and particles.
However, the intrinsic incompatibility inherent in quantum mechanics, coupled with the curse of dimensionality, renders the identification of optimal protocols and sensitivity bounds far richer and more challenging than in the single-parameter scenario.
For these reasons, advanced numerical techniques, including machine learning and variational approaches, may boost accuracy and enable real-time data processing and adaptive decision-making.
Finally, sensors exploiting quantum resources are generally highly sensitive to external noise and decoherence, which can significantly impair their performance. 
Implementing noise mitigation strategies, fault tolerance, and quantum error correction is therefore essential for maintaining high precision in applications.


{\bf General framework.}
A generic quantum probe state $\rhoop$ undergoes a quantum channel transformation that simultaneously encodes the values of $d$ parameters $\vect{\theta} = \{ \theta_1, ..., \theta_d \}^\top$ that we want to estimate.
Information about the unknown $\vect{\theta}$ is extracted from measurements performed on the transformed state $\rhoop_{\vect{\theta}}$.  
The detection process is generally described by a positive operator-valued measure (POVM), $\vect{\POVMop}=\{ \POVMop_1, \POVMop_2... \}^\top$, namely a set of non-negative operators ($\POVMop_k \geq 0$) satisfying the completeness relation $\sum_k \POVMop_k = \Eins$.
Each measurement outcome, labeled by $k$, occurs with probability $P(k \vert \vect{\theta}) = {\rm Tr}[\rhoop_{\vect{\theta}} E_k]$ according to the Born rule.
By repeating the measurement $m$ times, a sequence of independent outcomes $\vect{k} = \{ k_1, ..., k_m \}^\top$ is collected.
Finally, an estimator function, $\Theta_j(\vect{k})$ provides an estimate for each parameter $\theta_j$.
In the following, we assume locally-unbiased estimators, namely satisfying $\bar{\thest} = \sum_{\vect{k}} P(\vect{k} \vert \vect{\theta}) \thest(\vect{k}) = \vect{\theta}$ and $\vect{\nabla}_{\vect{\theta}} \bar{\thest} = \Eins_d$, where $\thest(\vect{k}) = \{ \Theta_1(\vect{k}), ..., \Theta_d(\vect{k})\}^\top$ and $P(k \vert \vect{\theta}) = \prod_{j=1}^m P(k_j \vert \vect{\theta})$ is the probability to observe the measurement sequence.
The aim of multiparameter quantum metrology is to identify the best combinations of probe states, POVMs, and estimators that enhance accuracy and precision, potentially incorporating optimal control protocols, ancillary subsystems, and/or joint measurements performed on multiple copies of $\rhoop_{\vect{\theta}}$.

\begin{figure}[ht!]
\includegraphics[width=1\columnwidth]{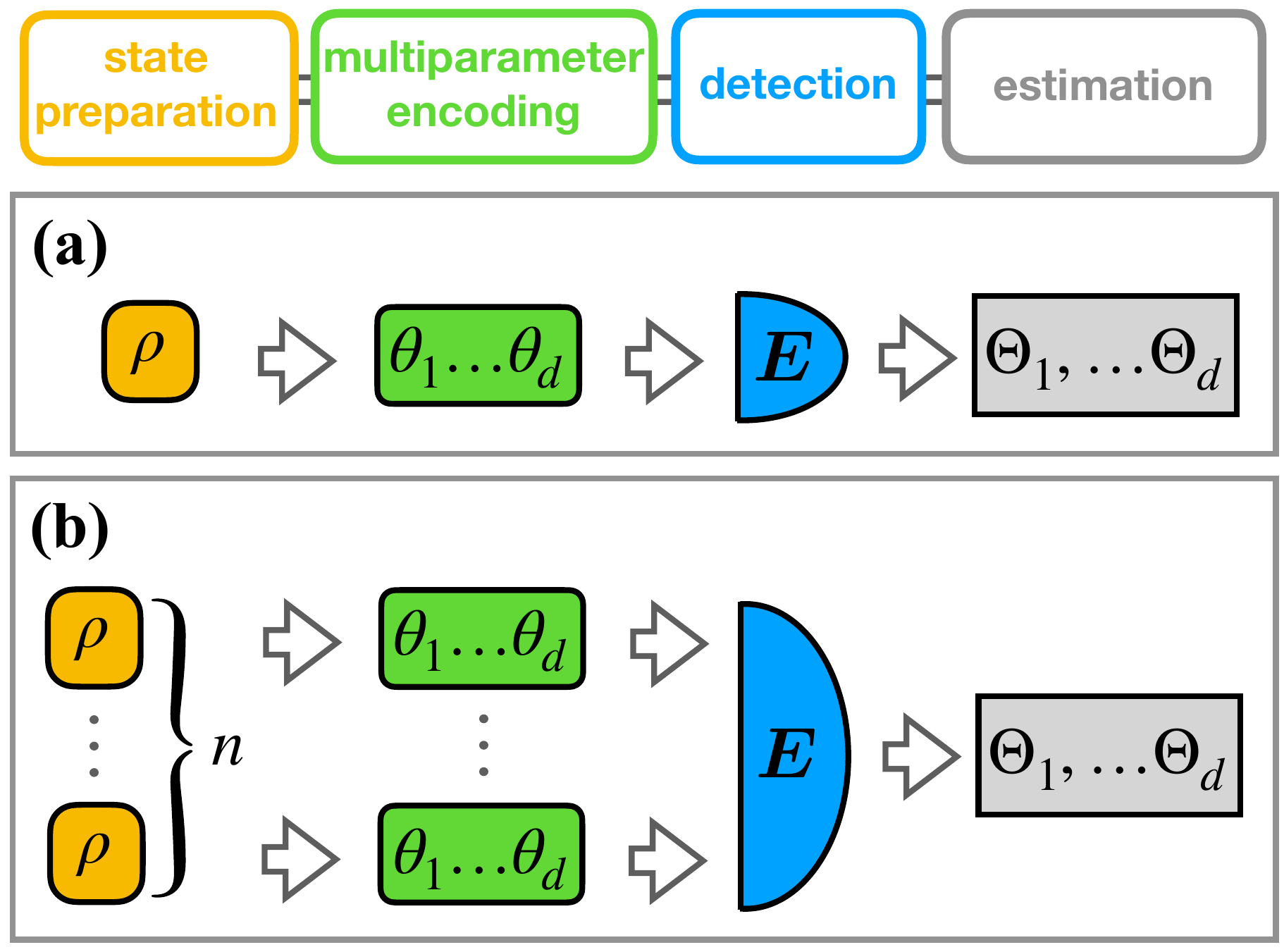}
\caption{{\bf Multiparameter estimation in the quantum setting.} 
Multiparameter estimation proceeds through a series of sequential steps: state preparation, simultaneous encoding of the several parameters described by a quantum channel, detection, and estimation.
In the simplest scenario (a), the protocol involves a single quantum probe state $\rho$: this is commonly referred to as a single-copy measurement.
In a more general scenario (b), joint measurements are performed on multiple ($n>1$) independent copies of the probe state, namely on $\rhotheta^{\otimes n}$.
In contrast to the single-parameter framework~\cite{GiovannettiNATPHOT2011,PezzeRMP2018}, joint measurements are typically optimal and necessary to saturate the ultimate quantum sensitivity bounds.
%
} \label{Figure1}
\end{figure}

{\bf Quantum bounds.} 
The estimation uncertainty is quantified by the $d \times d$ covariance matrix of estimators, 
\be \label{Eq.CovMat}
\vect{C}(\rhoop_{\vect{\theta}}, \vect{\POVMop}, \vect{\Theta}) 
= \sum_{\vect{k}} P(\vect{k} \vert \vect{\theta}) [\vect{\theta} - \vect{\thest}(\vect{k})][\vect{\theta} - \vect{\thest}(\vect{k})]^\top.
\ee
The diagonal element $\vect{C}_{jj}$ is the variance of $\Theta_j$, while the off-diagonal term $\vect{C}_{ij}$ provides statistical correlations between $\Theta_i$ and $\Theta_j$. 
Equation~(\ref{Eq.CovMat}) fulfills the chain of matrix inequalities~\cite{HelstromBOOK1976}
\be \label{matrixineq}
\vect{C}(\rhoop_{\vect{\theta}}, \vect{\POVMop}, \vect{\Theta}) \geq \frac{\vect{F}^{-1}(\rhoop_{\vect{\theta}}, \vect{\POVMop})}{m}\geq \frac{\vect{F}_Q^{-1}(\rhoop_{\vect{\theta}})}{m}.
\ee
The first inequality is the multiparameter Cram\'er-Rao bound (CRB). 
It is obtained by minimizing $\vect{C}$ over all possible unbiased estimators, where $\vect{F}(\rhoop_{\vect{\theta}}, \vect{\POVMop})$ is the Fisher information matrix (FIM), see Box.~\ref{box:CRB}.
The inequality can be saturated by optimal locally-unbiased estimators, such as the maximum
likelihood in the limit $m\gg 1$~\cite{KayBOOK1993}.
The second inequality is the multiparameter quantum Cram\'er-Rao bound (QCRB)~\cite{HelstromBOOK1976}, where $\vect{F}_Q(\rhoop_{\vect{\theta}})$ is the quantum Fisher information matrix (QFIM), see Box.~\ref{box:CRB}.
The meaning of Eq.~(\ref{matrixineq}) is that the variance  of any linear function of the parameters, $\vect{\nu}^\top \vect{\theta} = \sum_{j=1}^d \nu_j \theta_j$, fulfills 
\be \label{ineqC}
\Delta^2(\vect{\nu}^\top \vect{\theta}) =  \vect{\nu}^\top \vect{C} \vect{\nu} \geq 
\frac{\vect{\nu}^\top\vect{F}^{-1} \vect{\nu}}{m}\geq \frac{\vect{\nu}^\top \vect{F}_Q^{-1} \vect{\nu}}{m},
\ee
where the real vector $\vect{\nu} = \{\nu_1, ..., \nu_d\}^\top$ specifies the combination of interest.
The other combinations orthogonal to $\vect{\nu}$ play the role of nuisance parameters~\cite{SuzukiJPA2020} and generally affect $\Delta^2(\vect{\nu}^\top \vect{\theta})$.
Similarly to the single-parameter case ($d=1$)~\cite{BraunsteinPRL1994}, it can be proved that there is always an optimal $\vect{\nu}$-dependent POVM that saturates the last inequality in Eq.~(\ref{ineqC})~\cite{SuzukiJPA2020}.
Moreover, the eigenvector of the FIM (QFIM) corresponding to its largest eigenvalue determines the combination of the parameters that attains the lowest uncertainty for a given (optimal) POVM.

\begin{mybox}[label = {box:CRB}]{The quantum Fisher information matrix}
The FIM for the general scheme of Fig.~\ref{Figure1}(b), involving joint measurements performed on $n$ independent copies of $\rhoop_{\vect{\theta}}$, is
\be \label{Eq.F} \tag{A1}
\vect{F}(\rhoop_{\vect{\theta}}^{\otimes n}, \vect{\POVMop}) 
= \sum_{k} \frac{[\vect{\nabla}_{\vect{\theta}} P(k\vert \vect{\theta})][\vect{\nabla}_{\vect{\theta}} P(k\vert \vect{\theta})]^{\top}}{P(k\vert \vect{\theta})},
\ee
where the sum runs over all results of single-measurement results $k$, observed with probability $P(k \vert \vect{\theta}) = {\rm Tr}[\POVMop_k \rhoop_{\vect{\theta}}^{\otimes n}]$, and the POVM $\vect{\POVMop}$ describes measurements on $\rhoop_{\vect{\theta}}^{\otimes n}$.
The QFIM of the state $\rhotheta$ is~\cite{HelstromBOOK1976, LiuJPA2019}  
\be \label{QFIM} \tag{A2}
\vect{F}_Q(\rhoop_{\vect{\theta}}) \equiv \frac{1}{2}
{\rm Tr}[\rhoop_{\vect{\theta}} \{ \vect{\Lop}, \vect{\Lop}^\top\}] =
\Re( {\rm Tr}[\rhoop_{\vect{\theta}} \vect{\Lop} \vect{\Lop}^\top]),
\ee
where $\vect{\Lop} = \{L_{1}, ..., L_{d}\}^\top$ are Hermitian operators called symmetric logarithmic derivatives (SLDs) and defined, on the support of $\rhoop_{\vect{\theta}}$, by the relation $ \vect{\nabla}_{\vect{\theta}} \rhoop_{\vect{\theta}} = (\vect{\Lop} \rhoop_{\vect{\theta}} + \rhoop_{\vect{\theta}} \vect{\Lop})/2$.
Here and in the following, ${\rm Re}(x)$ and ${\rm Im}(x)$ indicate the real and imaginary part of $x$, respectively, while $[A,B]=AB-BA$ and $\{A,B\}=AB+BA$ are the commutator and anti-commutator of generic operators $A$ and $B$, respectively.
The QFIM is additive, $\vect{F}_Q(\rhoop_{\vect{\theta}}^{\otimes n}) = n \vect{F}_Q(\rhoop_{\vect{\theta}})$, and convex in the state, $\vect{F}_Q(\sum_\lambda q_\lambda \rhoop^{(\lambda)}_{\vect{\theta}})\leq \sum_\lambda q_\lambda \vect{F}_Q( \rhoop^{(\lambda)}_{\vect{\theta}})$~\cite{LiuJPA2019, GessnerPRL2018}.
While $\vect{F}(\rhoop^{\otimes n}_{\vect{\theta}}, \vect{\POVMop}) \leq n \vect{F_Q}(\rhoop_{\vect{\theta}})$ holds for every POVM, the equality cannot be always saturated, even asymptotically in $n$.

For pure states, Eq.~(\ref{QFIM}) simplifies to
\be  \tag{A3}
\vect{F}_Q(\vert \psi_{\vect{\theta}} \rangle) = 4~{\rm Cov}_{\ket{\psi}}(\vect{\mathcal{H}} \vect{\mathcal{H}}^{\top}), 
\ee
where $\ket{\psi_{\vect{\theta}}} = U_{\vect{\theta}} \ket{\psi}$, $U_{\vect{\theta}}$ the unitary evolution, $\vect{\mathcal{H}} = i U_{\vect{\theta}}^{\dag} \nabla_{\vect{\theta}} U_{\vect{\theta}} $ is a vector of the Hermitian operators generating the unitary encoding, and ${\rm Cov}_{\ket{\psi}}(A,B) = \bra{\psi} (AB+BA) \ket{\psi}/2 -\bra{\psi} A \ket{\psi} \bra{\psi} B \ket{\psi}$ is the symmetrized covariance between two generic operators.

The matrices $\vect{F}$ and $\vect{F}_Q$ are real, symmetric and positive semidefinite, and may be singular.
Consequently, $\vect{F}^{-1}$ and $\vect{F}_Q^{-1}$ are understood as the inverse on their respective support -- —that is, as the Moore-Penrose pseudo-inverses obtained by projecting onto the subspaces spanned by the eigenvectors associated with nonzero eigenvalues~\cite{ProctorPRL2018,GePRL2018}.
Physically, the FIM and QFIM become non-invertible when certain linear combinations of the parameters are either not independent or inaccessible.
In such cases, the Moore-Penrose pseudo-inverse effectively restricts the analysis to a smaller set of independent combinations.

\end{mybox}

In the multiparameter scenario, one may be interested in estimating several (not necessarily orthogonal) linear combinations of the parameters, $\vect{\nu}^\top_j \vect{\theta}$, with $j=1,2,...$.
Each combination is weighted by a positive factor $w_j>0$, reflecting the desired trade-off among the variances $\Delta^2(\vect{\nu}_j^\top \vect{\theta})$.
The figure of merit generalizes to $\sum_{j} w_j \Delta^2(\vect{\nu}_j^\top \vect{\theta})= {\rm Tr}[\vect{C}\vect{W}]$, where the $d\times d$ weight matrix $\vect{W} = \sum_{j} w_j \vect{\nu}_j \vect{\nu}_j^T$ is real, symmetric and positive definite.
In this case, finding the ultimate sensitivity bound -- referred to as the most informative bound (MIB) -- is highly nontrivial.
The MIB is defined as $\mathcal{B}_{\rm MI}(\rhoop_{\vect{\theta}}, \vect{W}) = \min_{\vect{\POVMop}} \mathcal{B}_{\rm CR}(\rhoop_{\vect{\theta}}, \vect{\POVMop},\vect{W})$, where $\mathcal{B}_{\rm CR}(\rhoop_{\vect{\theta}}, \vect{\POVMop},\vect{W}) = {\rm Tr}[\vect{W}\vect{F}^{-1}]$.
It corresponds to the minimum of $m {\rm Tr}[\vect{C} \vect{W}]$ over all possible locally-unbiased estimators and POVMs, and  is attainable for $m\gg 1$, at least.
Although the optimal POVMs achieving the MIB generally depend on the unknown $\vect{\theta}$, adaptive strategies can be employed to approach this bound when a sufficiently large number of copies of $\rhoop_{\vect{\theta}}$ is available~\cite{SuzukiJPA2020}.
If the optimal measurements associated to the different $\vect{\nu}_j$ commute, then the fundamental limitation is given by the QCRB, namely $\mathcal{B}_{\rm MI}(\rhoop_{\vect{\theta}}, \vect{W}) = \mathcal{B}_{\rm QCR}(\rhoop_{\vect{\theta}}, \vect{W}) = {\rm Tr}[\vect{W}\vect{F}_Q^{-1}]$.
However, if the optimal POVMs corresponding to different $\vect{\nu}_j$ do not commute, they are incompatible and cannot be performed at the same time. 
In this case, it is not possible to guarantee optimality for all $\Delta^2(\vect{\nu}_j^\top \vect{\theta})$ simultaneously, meaning that the QCRB may not be saturable for a general weight matrix $\vect{W}$.

The chain of inequalities
\beq \label{CW}
&& {\rm Tr}[\vect{C}(\rhoop_{\vect{\theta}}^{\otimes n}, \vect{\POVMop}, \vect{\Theta}) \vect{W}] 
 \geq \frac{\mathcal{B}_{\rm CR}(\rhoop_{\vect{\theta}}^{\otimes n}, \vect{\POVMop},\vect{W})}{m} \nonumber \\
&&  
\qquad
\geq \frac{\mathcal{B}_{\rm MI}(\rhoop_{\vect{\theta}}^{\otimes n}, \vect{W})}{m}
\geq \frac{\mathcal{B}_{\rm H}(\rhoop_{\vect{\theta}},\vect{W})}{n \, m} \geq \frac{\mathcal{B}_{\rm QCR}(\rhoop_{\vect{\theta}},\vect{W})}{n \, m}, \nonumber \\
\eeq
generalizes Eq.~(\ref{ineqC}) and can be derived, for any weight matrix $\vect{W}$, by considering the general protocol of Fig.~\ref{Figure1}(b).
There, $n$ independent copies of $\rhotheta$ are measured jointly. 
The protocol is eventually repeated $m$ times, thus involving $n\times m$ copies of $\rhotheta$, in total.
In the multiparameter scenario, joint measurements on $\rhoop_{\vect{\theta}}^{\otimes n}$ are generally more advantageous than performing $n$ independent measurements on single copies of $\rhoop_{\vect{\theta}}$: namely, $\mathcal{B}_{\rm MI}(\rhoop_{\vect{\theta}}^{\otimes n}, \vect{W}) \leq \mathcal{B}_{\rm MI}(\rhoop_{\vect{\theta}}, \vect{W})/n$~\cite{ChenPRL2022}.
For $n\to \infty$, the MIB, $\mathcal{B}_{\rm MI}(\rhoop_{\vect{\theta}}^{\otimes n}, \vect{W})$, converges to the Holevo bound (HB, sometimes also indicates as Holevo Cram\'er-Rao bound or Holevo-Nagaoka bound in the literature), $\mathcal{B}_{\rm H}(\rhoop_{\vect{\theta}}, \vect{W})/n$~\cite{HolevoBOOK1982, NagaokaBOOK}, see Box.~\ref{box:HCRB} for definitions and properties and Box~\ref{box:saturation} for an overview of saturation conditions~\cite{RafalPLA2020, AlbarelliPLA2020, SuzukiENTROPY2019}.
The experimental difficulty in implementing joint measurements~\cite{RocciaQST2018, ConlonNATPHYS2023} motivates the search for optimal POVMs -- only based on the knowledge of generic $\rhoop_{\vect{\theta}}$ and $\vect{W}$ --  that saturate the MIB for single-copy measurements: yet, it has been pointed out that this is one of the major open problems in quantum information theory~\cite{HorodeckiPRX2022}.
%
Recent progresses in this direction include computable~\cite{ConlonNPJQI2021} and tight~\cite{HayashiQUANTUM2023} bounds that are more informative than the HB.
Finally, $\mathcal{B}_{\rm MI}$ can be further minimized over all possible probe states $\rhoop$: this optimization is known only in the case of distributed sending with unitary parameter encoding, as it will be discussed below.

\begin{mybox}[label = {box:HCRB}]{The Holevo Bound}

To introduce the HB, let us first reformulate the QCRB as $\mathcal{B}_{\rm QCR}(\rhoop_{\vect{\theta}},\vect{W}) = \min_{\vect{X}} {\rm Tr}[\vect{W}\vect{Z}]$~\cite{HelstromBOOK1976}. 
Here, $\vect{Z}[\vect{X}]= {\rm Tr}[\rhoop_{\vect{\theta}}(\vect{X} - \vect{\theta}) (\vect{X} - \vect{\theta})^T]$ is a positive semi-definite $d\times d$ Hermitian matrix, and $\vect{X} = \sum_k \vect{\Theta}(k) \POVMop_k$ is an operator combining POVM and estimator, and satisfying the local-unbiasedness condition ${\rm Tr}[\rhoop_{\vect{\theta}} \vect{X}] = \vect{\theta}$ and ${\rm Tr}[\nabla_{\vect{\theta}} \rhoop_{\vect{\theta}} \vect{X}] = \Eins_d$.
The above minimization is formally achieved by $\vect{X}_Q = \vect{\theta} + \vect{F}_Q^{-1} \vect{L}$~\cite{HelstromBOOK1976}.
Yet, since $\vect{W}$ is real and symmetric, taking the trace ${\rm Tr}[\vect{W}\vect{Z}[\vect{X}]]$ causes the positive contribution due to the imaginary part of $\vect{Z}$ -- related to the possible non-commutativity $[X_i,X_j]\neq 0$ of POVMs -- to be lost, resulting in an overly optimistic bound.
Holevo thus proposed to generalize the QCRB as $\mathcal{B}_{\rm H}(\rhoop_{\vect{\theta}},\vect{W}) = \min_{\vect{X}, \vect{V}} \{ {\rm Tr}[\vect{W}\vect{V}] \vert \vect{V} \geq \vect{Z}[\vect{X}] \}$~\cite{HolevoBOOK1982}, where $\vect{V}$ is a real matrix: thanks to the constraint $\vect{V} \geq \vect{Z}[\vect{X}]$, computing the bound $\mathcal{B}_{\rm H}$ involves both the real and the imaginary parts of $\vect{Z}$.
As pointed out by Nagaoka~\cite{NagaokaBOOK},
the minimization over $\vect{V}$ can be performed analytically, giving
\begin{align} \label{HCR} 
   & \mathcal{B}_{\rm H}(\rhoop_{\vect{\theta}},\vect{W}) = \min_{\vect{X}} \Big\{ {\rm tr}[\vect{W} \Re\vect{Z}[\vect{X}]] + \nonumber \\
& \qquad \qquad  \qquad  + \vert\vert\sqrt{\vect{W}} \Im\vect{Z}[\vect{X}]  \sqrt{\vect{W}} \vert\vert_1 \Big\}, \tag{B1} 
\end{align}
where $\vert\vert A \vert\vert_1 = {\rm Tr}[\sqrt{A^\dag A}]$ is the trace norm.
The minimum in Eq.~(\ref{HCR}) always exists for finite dimensional systems~\cite{SuzukiJPA2020}.
Taking $\vect{X} = \vect{X}_Q$ shows that $\mathcal{B}_{\rm H}(\rhoop_{\vect{\theta}},\vect{W}) \geq \mathcal{B}_{\rm QCR}(\rhoop_{\vect{\theta}},\vect{W})$.

While the QCRB can be calculated directly from the quantum state $\rho_{\vect{\theta}}$, there is no general closed form for Eq.~(\ref{HCR}).
The additivity property $\mathcal{B}_{\rm H}(\rhoop_{\vect{\theta}}^{\otimes n},\vect{W}) = \mathcal{B}_{\rm H}(\rhoop_{\vect{\theta}},\vect{W})/n$ \cite{HayashiJMP2008}, is highly non-trivial.
It is used in Eq.~(\ref{CW}) and implies that it is sufficient to calculate the HB for a single copy of $\rhoop_{\vect{\theta}}$.
General expressions for the HB are available in the case of two parameters ($d=2$)~\cite{SuzukiJMP2016, MatsumotoJPA2002, BradshawPRA2018, SidhuPRX2021}.
For a specific class of problems known as D-invariant~\cite{HolevoBOOK1982} -- that includes the tomography of finite-dimensional quantum systems, see Refs.~\cite{SuzukiJMP2016, SuzukiENTROPY2019} -- the minimization in Eq.~(\ref{HCR}) is achieved by $\vect{X} = \vect{X}_Q$. 
More generally, the evaluation of the HB for finite dimensional systems and arbitrary number of parameters can be recast as a linear semi-definite program~\cite{AlbarelliPRL2019}, which is feasible for numerical computation.
Numerical methods to derive upper and lower bounds to the HCRB have been discussed in Ref. \cite{SidhuPRX2021}.

\end{mybox}

\begin{mybox}[label = {box:saturation}]{Saturation of the quantum bounds}

For a generic mixed state $\rhoop_{\vect{\theta}}$ and weight matrix $\vect{W}$, the HB asymptotically approaches the MIB when joint measurements are performed on the multi-copy probe $\rhoop^{\otimes n}_{\vect{\theta}}$: the saturation is guaranteed in the limit $n\to \infty$~\cite{KahnCMM2009, YamagataANNSTAT2013,YangCMM2019}.
The QCRB coincides with the HB for every $\vect{W}$ if and only if the so-called weak commutativity condition
\be \label{saturationGENERAL} \tag{C1}
\vect{G}_Q(\rhoop_{\vect{\theta}}) \equiv \frac{1}{2i}{\rm tr}[\rhoop_{\vect{\theta}}[\vect{\Lop}, \vect{\Lop}^\top]] = \Im({\rm Tr}[\rhoop_{\vect{\theta}}\vect{\Lop} \vect{\Lop}^\top]) = 0,
\ee
holds~\cite{MatsumotoJPA2002, RagyPRA2016}.
If the condition (\ref{saturationGENERAL}) is met, the QCRB saturates the MIB when collective measurements are performed on $\rhoop^{\otimes n}_{\vect{\theta}}$, in the limit $n\to \infty$.
A fundamental limitation, known as gap persistence theorem~\cite{ConlonARXIV}, states that if the HB or QCRB cannot be attained with a single copy of the probe ($n=1$), then these bounds remain unattainable by any measurement of $\rhoop_{\vect{\theta}}^{\otimes n}$, for any finite $n$.

The proof that $\mathcal{B}_{\rm CR}(\rhoop_{\vect{\theta}}, \vect{W}) \leq \mathcal{B}_{\rm QCR}(\rhoop_{\vect{\theta}}, \vect{W})$~\cite{HelstromBOOK1976} and $\vect{F}(\rhoop_{\vect{\theta}}, \vect{\POVMop}) \leq  \vect{F_Q}(\rhoop_{\vect{\theta}})$~\cite{YangPRA2019} are based on Cauchy-Schwartz inequalities: the necessary and sufficient conditions for the saturation are known~\cite{HelstromBOOK1976,YangPRA2019,SuzukiJPA2020} but they are highly nontrivial and cannot be expressed solely in terms of the state $\rhoop_{\vect{\theta}}$.
It has been also pointed out~\cite{ConlonARXIV, ConlonARXIV2024} that the QCRB can be saturated, for any $n$, if and only if generalized SLD operators commute on an extended Hilbert space.
Interestingly, the so-called partial commutativity $\Pi [\vect{L}_{\vect{\theta}}, \vect{L}_{\vect{\theta}}^T] \Pi = 0$, where $\Pi$ is the projector on the support of $\rhoop_{\vect{\theta}}$, is necessary (but not sufficient~\cite{ConlonARXIV2024}) for the saturation of the QCRB~\cite{YangPRA2019}.
Generalizations of such a necessary condition in the case of joint measurements on $\rhoop_{\vect{\theta}}^{\otimes n}$ have been provided in Ref.~\cite{ChenPRL2022}.
For mixed states of full rank, we have that $\Pi$ span the full Hilbert space ($\Pi = \Eins$) and therefore full commutativity $[\vect{L}_{\vect{\theta}}, \vect{L}_{\vect{\theta}}^T] = 0$ becomes necessary and sufficient: the optimal POVM being given by the projection over the eigenstates of each SLD, $L_j$.  

For pure states, the HB always saturates the MIB without requiring joint measurement (namely, for $n=1$)~\cite{MatsumotoJPA2002}.
Equation~(\ref{saturationGENERAL}) rewrites as 
\be \label{saturationPURESTATES} \tag{C2}
\frac{1}{2} \bra{\psi} [\vect{\Hop}, \vect{\Hop}^\top] \ket{\psi} = 
\Im(\bra{\psi}\vect{\mathcal{H}} \vect{\mathcal{H}}^\top \ket{\psi})=0,
\ee
where $\vect{\mathcal{H}}$ is defined in Box.~\ref{box:CRB}.
Equation (\ref{saturationPURESTATES}) is equivalent to the weak commutativity condition Eq.~(\ref{saturationGENERAL}) and is necessary and sufficient for the existence of a POVM such that $\vect{F}(\ket{\psi_{\vect{\theta}}}, \vect{\POVMop}) = \vect{F_Q}(\ket{\psi_{\vect{\theta}}})$~\cite{MatsumotoJPA2002}.
In this case, necessary and sufficient conditions for optimal projective measurements have been derived in Ref.~\cite{PezzePRL2017}, including receipts to construct such projectors.

Finally, we clarify that the above saturation conditions consider the case of arbitrary $\vect{W}$. 
For instance, the HB coincides with the QCRB for any weight matrix of unit rank, namely $\vect{W}=\vect{\nu}\vect{\nu}^\top$, $\mathcal{B}_{\rm H}(\rhoop_{\vect{\theta}},\vect{\nu}\vect{\nu}^\top) = \mathcal{B}_{\rm QCR}(\rhoop_{\vect{\theta}},\vect{\nu}\vect{\nu}^\top)$~\cite{RafalPLA2020, SuzukiJPA2020, TsangPRX2020}, achievable with measurements on single copies of $\rhoop_{\vect{\theta}}$.

\end{mybox}

\begin{figure*}[ht!]
\includegraphics[width=0.93\textwidth]{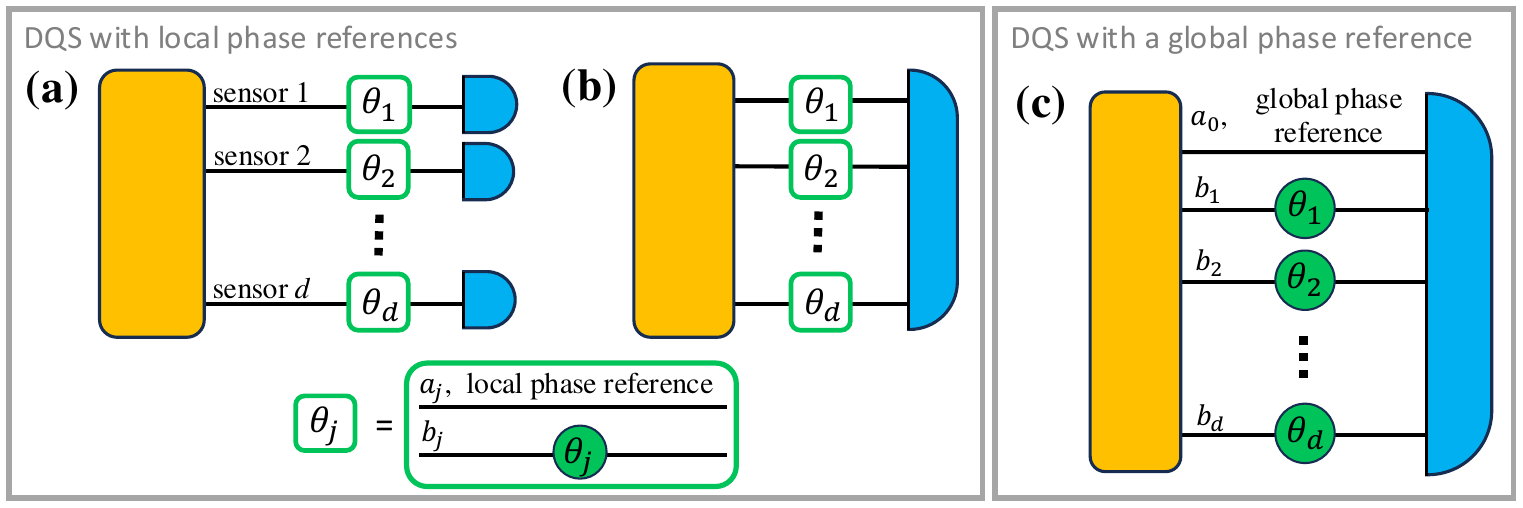}
\caption{{\bf Models of DQS}. 
Panels (a) and (b) show DQS schemes with local phase references. 
The two schemes differs by local (a) or global (b) measurements, represented by the blue semicircle(s).
Green boxes are local sensors.
In the simplest case, each self-referenced sensor is a two-mode device (inset) with $\theta_j$ being the relative phase shifts between mode $a_j$ and $b_j$, one mode providing the local phase reference. 
The scheme (a) shows a general configuration of DQS with local phase references.
Here, the parameter $\theta_j$ is the relative phase shifts between modes $a_0$ and the mode $b_j$ of the $j$th local sensor.
In this scheme, global measurements are necessary.
In all panels, the yellow box represents schematically the probe state preparation.
}
\label{Fig1}
\end{figure*}

{\bf Incompatibilities.}
The ratio between the HB and the QCRB is~\cite{CarolloJSM2019}
\be \label{Rineq}
1 \leq  \frac{\mathcal{B}_{\rm H}(\rhoop_{\vect{\theta}},\vect{W})}{\mathcal{B}_{\rm QCR}(\rhoop_{\vect{\theta}},\vect{W})} \leq 1 + \mathcal{R} \leq 2.
\ee
Here $\mathcal{R} = \vert \vert i \vect{F}_Q^{-1} \vect{G}_Q \vert\vert_{\infty}$, the norm $\vert \vert \cdot \vert\vert_{\infty}$ indicates the largest eigenvalue of a matrix, and $\vect{G}_Q$, defined in Eq.~(\ref{saturationGENERAL}), can be non-zero only in a multiparameter scenario.
Interestingly, $0 \leq \mathcal{R} \leq 1$, with $\mathcal{R}=0$ if and only if $\vect{G}_Q=0$~\cite{CarolloJSM2019}. 
The quantity $\mathcal{R}$ has been thus proposed as a figure of merit of {\it measurement incompatibility}~\cite{CarolloJSM2019, BelliardoNJP2021,RazavianENTROPY2020}, see also Refs.~\cite{KullJPA2020,LuPRL2021} for an alternative approaches based on uncertainty relations.
In the case of maximal incompatibility, $\mathcal{R}=1$, the HB is at most twice the QCRB~\cite{CarolloJSM2019, TsangPRX2020}.
This factor two can be qualitatively understood within the theory of quantum local asymptotic normality~\cite{KahnCMM2009}, see Ref.~\cite{RafalPLA2020} for a review.
According to this framework, in the limit $n\to \infty$, any quantum statistical model becomes equivalent to a Gaussian shift model and measurement incompatibility can be understood in terms of effective position and momentum displacement estimations.

For a given probe, certain combinations of the parameters cannot be estimated with the same sensitivity as others: a problem generally indicated as {\it probe incompatibility}.
Assuming that the QFIM can be achieved, probe incompatibility is linked to the widely distributed spectrum of the QFIM.
This is expressed by the chain of inequalities (here $n=1$, for simplicity)
\be \label{waekCRB}
{\rm Tr}[\vect{W}\vect{C}] \geq \sum_j w_j \frac{\vect{\nu}_j^{\top} \vect{F}_Q^{-1} \vect{\nu}_j}{m} \geq \sum_j w_j \frac{(\vect{\nu}^\top_j\vect{\nu}_j)^2}{m \, \vect{\nu}^\top_j \vect{F}_Q \vect{\nu}_j}.
\ee
The right-hand side provides a weak form of the QCRB that is often considered since it avoids computing the inverse of the QFIM~\cite{GePRL2018, GessnerPRL2018}: the bound is saturated, for arbitrary $w_j$, if and only if $\vect{\nu}_j$ is an eigenstate of the QFIM~\cite{MalitestaArXiv}.
For weight matrices of full rank, Eq.~(\ref{waekCRB}) can be further bounded by $d^2\big[\sum_{j=1}^d \tfrac{m \vect{\nu}_j^{\top} \vect{F}_Q^{-1} \vect{\nu}_j}{w_j (\vect{\nu}^\top_j\vect{\nu}_j)^2}\big]^{-1}$ with equality if and only if $\vect{W}_{\rm opt} = \lambda \vect{F}_Q$.
For a given $\vect{F}_Q$, $\vect{W}_{\rm opt}$ provides the optimal choice of weight matrix, ${\rm Tr}[\vect{W}\vect{C}] \geq {\rm Tr}[\vect{W}_{\rm opt}\vect{C}] = \lambda d/m$.
Viceversa, taking an identity weight matrix, $\vect{W} = \Eins_d$, corresponding to ${\rm Tr}[\vect{C}] = \sum_{j=1}^d \Delta^2 \theta_j$ being the sum of single-parameter variances, optimal states $\rhotheta$ have a diagonal QFIM~\cite{RagyPRA2016,NicholsPRA2018,SuzukiJPA2020}. 

{\bf Distributed quantum sensing.}
Distributed quantum sensing (DQS) consists of a network of $d$ spatially-delocalized sensors.
In the ideal scenario, the parameter-encoding transformations is $\bigotimes_{j=1}^d e^{-i\theta_j H_j}$, where the local Hamiltonians $H_j$ modeling each sensor commute: $[\hat{H}_i, \hat{H}_j] =0$,
for all $i,j=1, ...,d$.
This working condition has two advantages: it identifies the Hamiltonians as the generators of the local unitary evolution, $H_j=\mathcal{H}_j$, see Box.~\ref{box:CRB}, and guarantees the saturation of the QCRB for pure states and with single-copy measurements, according to Eq.~(\ref{saturationPURESTATES}). 

%
By distributing an entangled state among different sensing nodes, DQS enables the estimation of local parameters $\theta_1, ..., \theta_d$, or their linear combinations, with a sensitivity that surpasses what is achievable using unentangled states.
DQS has been mainly explored in photonic platforms, particularly estimating multiple phase shifts across distinct optical paths.
The generalization to atomic platforms includes applications such as differential interferometry~\cite{CorgierQUANTUM2023} and gradient magnetometry~\cite{ApellanizPRA2018}.
Two main DQS architectures have been studied in the literature, differing in their phase reference configurations, as illustrated in Fig.~\ref{Fig1}.

\begin{figure}[b!]
\includegraphics[width=1\columnwidth]{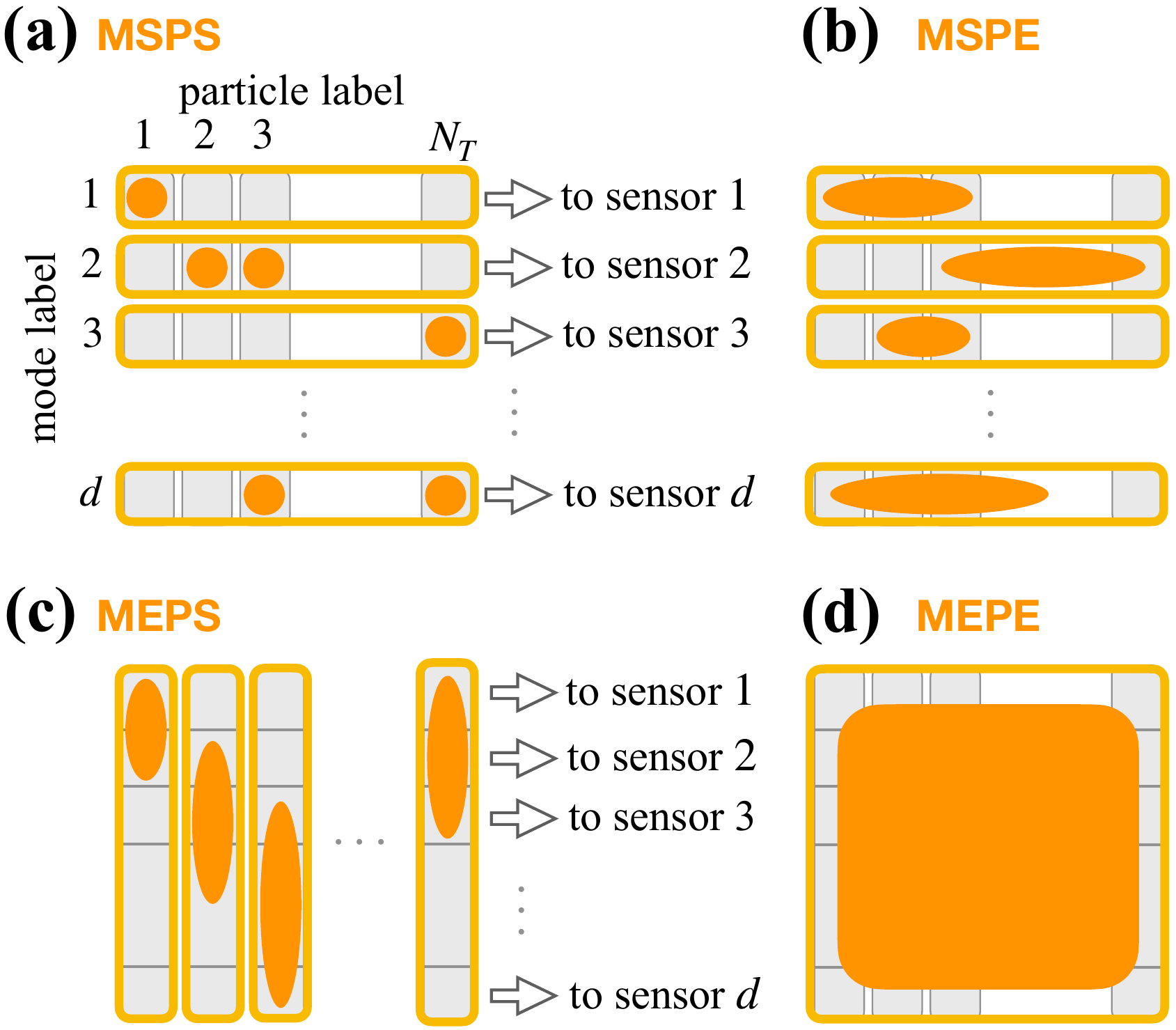}
\caption{{\bf Quantum resources in DQS with local phase references.} 
A DQS probe state of $N_T$ particles is distributed among $d$ modes.
The horizontal (vertical) lines represent the particle (mode) partition of the quantum state.
The probe state can be prepared as 
(a) mode and particle separable (MSPS), 
(b) mode separable and particle entangled (MSPE),
(c) mode entangled and particle separable (MEPS), or
(d) mode and particle entangled (MEPE). 
Mode (particle) entanglement is schematically illustrated by broad vertical (horizontal)  ensembles encompassing multiple labels.
}
\label{Fig2}
\end{figure}

{\it DQS with local phase references.}
In this configuration, each sensor performs a relative phase measurement~\cite{GePRL2018, MalitestaArXiv, ZhuangPRA2018,  PezzeARXIV2024}, see Fig.~\ref{Fig1}(a-b).
The $j$th sensor comprises two modes, $a_j$ and $b_j$, with $H_j = (N_j^{(a)} - N_j^{(b)})/2$, $N_j^{(a)}$ and $N_j^{(b)}$ are number of particles operators, and $\theta_j$ is the relative phase shift among the two modes.
Optimal measurements schemes can be local at each sensor, Fig~\ref{Fig1}(a), or global, Fig~\ref{Fig1}(b), requiring the recombination of the different sensing modes after phase encoding. 
Depending on the properties of the probe state, we recognize four possible scenarios, see Fig.~\ref{Fig2}, contingent on the utilization of mode- and/or particle-entanglement~\cite{GessnerPRL2018,LiuNATPHOT2020}.
In the following, the four possibilities are compared by using the same total number of particles, $N_T$, and, for simplicity, by taking $\Delta^2 (\vect{\nu}^\top_{\rm ave} \vect{\theta})$ as the figure of merit, where $\vect{\nu}_{\rm ave} = \{1, ..., 1\}^\top/d$ and $\vect{\nu}^\top_{\rm ave} \vect{\theta} = \tfrac{1}{d}\sum_{j=1}^d \theta_j$ is the average of the $d$ phases.
This combination of parameters provides the maximum sensitivity enhancement offered by entanglement.
In this case, all protocols of Fig.~\ref{Fig2} are optimized by using the same number of particles $N=N_T/d$ (assumed integer) in each sensor. 

When the probe state is mode-separable (MS), the $d$ sensors operate independently, see Fig.~\ref{Fig2}(a) and (b).
In this case, the QFIM is diagonal: the sensitivity bounds follow the single-parameter case~\cite{GiovannettiNATPHOT2011, PezzeRMP2018} and can be saturated with local measurements at each sensor.
The highest sensitivity of the mode-separable-particle-separable (MSPS) strategy of Fig.~\ref{Fig2}(a) is the standard quantum limit (SQL)
\be\label{SQL}
\Delta^2 (\vect{\nu}^\top_{\rm ave} \vect{\theta})_{\rm MSPS} 
= \frac{1}{m N d} = \frac{1}{m N_T},
\ee
achieved with 
\be \label{PsiPsMs}
\ket{\psi_{\rm MSPS}} = \bigotimes_{j=1}^d \bigg( \frac{\ket{1,0}_j+\ket{0,1}_j}{\sqrt{2}} \bigg)^{\otimes N}.
\ee
Here, $\ket{N,0}_j$ ($\ket{0,N}_j$) is a state of $N$ ($0$) particles -- assumed integer -- in the mode $a_j$ ($b_j$) of the $j$th sensor and $0$ ($N$) particles in the other mode, see Fig.~\ref{Fig1}(a-b).
Particle entanglement, Fig.~\ref{Fig2}(b), is necessary to overcome the SQL for the estimation of each $\theta_j$~\cite{PezzePRL2009}.
The maximum sensitivity achievable with a mode-separable-particle-entangled (MSPE) probe, is 
\be \label{Eq.SHL}
\Delta^2 (\vect{\nu}_{\rm ave}^\top\vect{\theta})_{\rm MSPE} =
\frac{1}{m N^2 d} = \frac{d}{m N_T^2},
\ee
obtained when using a product of NOON states
\be \label{PsiPeMs}
\ket{\psi_{\rm MSPE}} = \bigotimes_{j=1}^d \frac{\ket{N,0}_j+\ket{0,N}_j}{\sqrt{2}}.
\ee
Equation (\ref{Eq.SHL}) overcomes the SQL by a factor $N$, equal to the number of particles in each sensor~\cite{PezzePRL2009, GiovannettiNATPHOT2011, PezzeRMP2018}.

A mode-entangled (ME) probe establishes quantum correlations between the modes of the different sensors.
The possible benefit of ME depends on whether the figure of merit ${\rm Tr}[\vect{W \vect{F}_Q^{-1}}]$ takes advantage of the off-diagonal elements of the QFIM. 
If the weight matrix $\vect{W}$ is diagonal, then MS states can enable an estimation uncertainty that is at least as small as the one that can be achieved with ME states~\cite{ProctorPRL2018}, as a direct consequence of the probe incompatibility discussed previously.
In contrast, ME can play a relevant role to reduce the uncertainty for the estimation of linear combinations of parameters, namely for $\vect{W} = \vect{\nu} \vect{\nu}^T$~\cite{ProctorPRL2018, GePRL2018}.
In general, the saturation of the QCRB with ME states requires global measurements, as in Fig.~\ref{Fig1}(b).
Overall, the QFIM is a relevant figure of merit to study the interplay of useful mode- and particle-entanglement in DQS~\cite{GessnerNATCOMM2020}.
The mode-entangled-particle-separable (MEPS) strategy of Fig.~\ref{Fig2}(c) is obtained by distributing independent single particles over the $d$ sensors. 
The corresponding sensitivity, $\Delta^2 (\vect{\nu}^\top\vect{\theta})_{\rm MEPS}$ reaches the SQL, at best~\cite{GessnerPRL2018}.
To overcome Eq.~(\ref{SQL}), particle entanglement is necessary.
The ultimate bound when using mode-entangled-particle-entangled (MEPE) states is~\cite{GessnerPRL2018,ProctorPRL2018}
\be \label{Eq.EHL}
\Delta^2 (\vect{\nu}_{\rm ave}^\top\vect{\theta})_{\rm MEPE} =
\frac{1}{m N^2 d^2} = \frac{1}{m N_T^2},
\ee
which can be achieved with the multi-mode NOON-like state~\cite{GessnerPRL2018, ProctorPRL2018}
\be \label{MePe}
\ket{\psi_{\rm MEPE}} = \frac{\otimes_{j=1}^d\ket{N,0}_j+\otimes_{j=1}^d\ket{0,N}_j}{\sqrt{2}}.
\ee
We remark that Eq.~(\ref{MePe}) is only sensitive to the specific combination of the parameters $\vect{\nu}_{\rm ave}^\top\vect{\theta}$
: the different $\theta_j$ cannot be estimated separately and DQS reduces effectively to a single-parameter estimation problem (the QFIM is of rank-1).
Equation (\ref{Eq.EHL}) is a factor $d$ smaller than Eq.~(\ref{Eq.SHL}). 
This sensitivity enhancement is a direct consequence of the larger number of entangled particles~\cite{LiuNATPHOT2020}: in the state Eq.~(\ref{MePe}) all $N_T$ particles are entangled, while only $N = N_T/d$ particles are entangled in the state Eq.~(\ref{PsiPeMs}).

The above results can be generalized to a arbitrary linear combinations $\vect{\nu}^\top\vect{\theta}$ upon optimizing the NOON-like state as in Eq.~(\ref{MePe}) and the distribution of particles in each sensor.
For example, the state $(\ket{N,0}_1\ket{0,N}_2+\ket{0,N}_1\ket{N,0}_2)/\sqrt{2}$ can be used to estimate the difference $\theta_1-\theta_2$ with sensitivity given by Eq.~(\ref{Eq.EHL}).
For a generic $\vect{\nu}$, the maximum gain of the MEPE strategy over the MSPE one is $\mathcal{G}(\vect{\nu}) = \Delta^2 (\vect{\nu}^\top\vect{\theta})_{\rm MSPE}/\Delta^2 (\vect{\nu}^\top\vect{\theta})_{\rm MEPE} = \lVert \vect{\nu} \rVert^2_{2/3}/\lVert \vect{\nu} \rVert^2_1$~\cite{ProctorPRL2018}, where $\lVert \vect{\nu} \rVert_\gamma=(\sum_{j=1}^d \vert \nu_j \vert^{\gamma})^{1/\gamma}$.
The gain $\mathcal{G}$ ranges from one, when estimating a single parameter (e.g. $\vect{\nu} = \{1, 0, ..., 0\}^\top$), to a maximum $\mathcal{G}=d$, achieved for $\vect{\nu} = \vect{\nu}_{\rm ave}$~\cite{ProctorPRL2018, GessnerPRL2018, GePRL2018}.
Finally, a DQS protocol to estimate arbitrary analytical functions of $\vect{\theta}$ using ME states has been discussed in Ref.~\cite{QianPRA2019}.

DQS with NOON-like (or GHZ-like, in the case of distinguishable qubits) states has been explored experimentally.
The case of two qubits ($N=2$ and $d=2$) prepared in a Bell state [analogous to Eq.~(\ref{MePe})] has been realized with photons~\cite{ZhaoPRX2021} and trapped Strontium ions~\cite {NicholNATURE2022}.
The former experiment has demonstrated distributed sensing with unconditional (without post-selection) sensitivity overcoming the SQL by 0.92 dB. 
The experiment of Ref.~\cite{NicholNATURE2022} has realized a quantum network of entangled atomic clocks, reporting a sensitivity beyond the SQL for the estimation of the frequency difference between two clocks.
The experiment~\cite{LiuNATPHOT2020} has investigated DQS using the state Eq.~(\ref{MePe}) for $d=3$ sensors and $N_T=6$ photons in total.
This experiment has demonstrated 2.7 dB of gain over the SQL for the estimation of the average phase $(\theta_1+\theta_2+\theta_3)/3$, also combined with a multi-round protocol.
It has been also emphasized~\cite{KimNATCOMM2024} that the MEPE state of Eq.~(\ref{MePe}) requires $N\geq d$ particles to estimate an equally-weighted sum of parameters.
Alternative quantum states that circumvent this difficulty have been proposed and experimentally implemented with $d=4$ and $N=2$ photonic qubits~\cite{KimNATCOMM2024}, reaching 2.2 dB sensitivity enhancement over the SQL.
While most proof-of-principle DSQ experiments are confined to laboratory-scale setups, Refs.~\cite{ZhaoPRX2021, KimNATCOMM2024} have demonstrated phase shift estimations over fiber distances exceeding a kilometer.

\begin{mybox}[label = {box:MZI}]{Mach-Zehnder sensor network} 

An array of MZIs~\cite{GessnerNATCOMM2020, MaliaNATURE2022, MalitestaArXiv, PezzeARXIV2024}, realizes a DQS scheme with local phase references.
In the scheme
\begin{center}
\includegraphics[width=1\columnwidth]{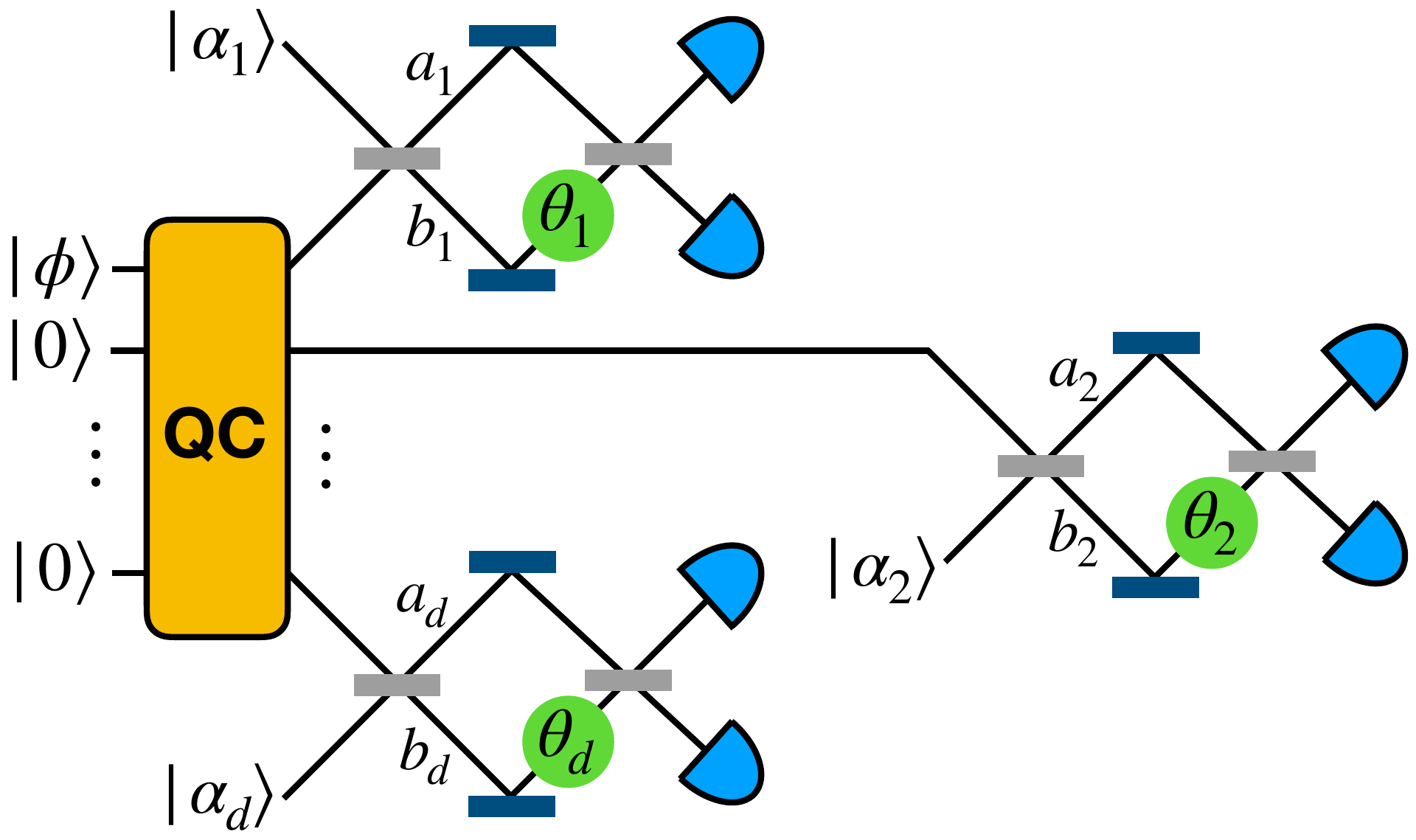}
\end{center}
a generic single-mode non-classical state $\ket{\phi}$ (e.g. a squeezed or a Fock state) is distributed in the network by using a $d$-mode QC, generating mode-entanglement.  
One output of the QC is used as input of the $j$th MZI of the network, the other input being a coherent state $\ket{\alpha_j}$~\cite{MalitestaArXiv, PezzeARXIV2024}.
The scheme can be optimized with respect to $\ket{\phi}$, the splitting performed by the QC, and the intensities of the coherent states.
It can reach sub-SQL sensitivities $\Delta^2 (\vect{\nu}^\top \vect{\theta})<1/(m N_T)$ for arbitrary $\vect{\nu}$, by using local measurements at the outputs of each MZI~\cite{PezzeARXIV2024}, as in Fig.~\ref{Fig1}(a).
Here, $N_T$ here is average number of particles in the  state $\ket{\phi}$ plus the total intensity of the $d$ coherent states.
In contrast, $d$ independent MZIs can achieve a $\Delta^2 (\vect{\nu}^\top \vect{\theta})$ overcoming the SQL for a generic $\vect{\nu}$ only at the price of operating $d$ non-classical states simultaneously.
The ME scheme offers significant advantages by reducing resource overhead, while enhancing estimation sensitivity.  
Furthermore, when $\ket{\phi}$ is the squeezed-vacuum, the ME scheme can achieve the sensitivity given by Eq.~(\ref{Eq.EHL}). 
In the limit of large intensity of each coherent state, each MZI in the sensor network performs a quadrature displacement~\cite{ZhuangPRA2018, OhPRR2020}. 
This limit, has been realized experimentally in a radio-frequency photonic sensor network~\cite{XiaPRL2020} with three sensing nodes. 
This experiment has achieved a sensitivity 3.2 dB below the SQL and has demonstrated the connection between ME structure and quantum noise reduction for various $\vect{\nu}$.

\end{mybox}

Since the NOON/GHZ-like states as in Eq.~(\ref{MePe}) are fragile and difficult to create with a large number of particles, several works have focused on DQS schemes exploiting more robust  states~\cite{GePRL2018,GessnerNATCOMM2020}.  
A practical strategy to generate useful ME states for DQS applications is to split a non-classical state by using a linear beam splitting network~\cite{ClementsOPTICA2016}, also indicated as quantum circuit (QC) or tritter (quarter) for $d=3$ ($d=4$).
As an example, Box.~\ref{box:MZI} shows a sensing scheme involving a network of Mach-Zehnder interferometers (MZIs) that uses a single non-classical state distributed by a QC.
Such a splitting may introduce additional noise and losses, which limits scalability and performance of the ME sensor network. 
Furthermore the QC must be optimized in order to minimize the uncertainty $\Delta^2(\vect{\nu}^\top \vect{\theta})$, for a given $\vect{\nu}$.
The experiment~\cite{MaliaNATURE2022} has investigated the case of two atomic interferometers where a suitable four-mode spin-squeezed state generated by quantum non-demolition measurement is used to estimate the differential phase shift $\theta_1 - \theta_2$ with a sensitivity overcoming the SQL by 11.6 dB. 
In addition, DQS using bright two-mode squeezed light in a SU(1,1) interferometer has been reported in Ref.~\cite{HongARXIV2024}, estimating an optimal linear combination of $d=2$ local phase shifts with sensitivities 1.7 dB below the SQL and using local homodyne detection.

{\it DQS with a global phase reference.}
The scheme of Fig.~\ref{Fig1}(c) consists of $d+1$ modes and is sensitive to the $d$ relative phase shifts $\theta_j$ with respect to a common phase reference.
The generator of $j$th phase encoding is the number of particle operator $H_j = N_j$ [represented schematically by the green circle in Fig.~\ref{Fig1}(c)].
The scheme can be understood as a multimode MZI for the estimation of the $d$ phases, where a global measurement recombining the sensing modes is necessary.
The scheme, first proposed in Ref.~\cite{HumphreysPRL2013} in the framework of discrete variables, naturally involves ME and is conveniently described in the qudit formalism~\cite{CiampiniSCIREP2016}, where each qudit is a single particle distributed among the $d+1$-modes. 
The potential advantage of DQS can emerge in the simultaneous estimation of each phase individually and is captured by the figure of merit ${\rm Tr}[\vect{C}] = \sum_{j=1}^d\Delta^2 \theta_j$, given by the sum of estimation variances.
When using the generalized NOON-like state~\cite{HumphreysPRL2013} 
\be  \label{psidatta}
\ket{\psi(N_T)} =  
\frac{\ket{N_T,0,...,0}}{\sqrt{1+\sqrt{d}}} + \frac{\ket{0,N_T, 0,...,0} + ... + \ket{0,...,0,N_T}}{\sqrt{d+\sqrt{d}} }
\ee
it is possible to reach
\be \label{MEPEdatta}
{\rm Tr}[\vect{C}]_{\ket{\psi(N_T)}} = \frac{d(\sqrt{d}+1)^2}{4N_T^2m},
\ee
where $\ket{N_0,N_1,..,N_d}$ is a Fock state of $N_j$ particles in mode $j=0, ..., d$, created before phase encoding in Fig.~\ref{Fig2}(c).
Equation (\ref{MEPEdatta}) is a factor $N_T$ smaller than the ${\rm Tr}[\vect{C}]_{\ket{\psi(1)}^{\otimes N_T}} = d(\sqrt{d}+1)^2/(4 N_T m)$ achieved with the MS state of $N_T$ qudits, namely $\ket{\psi(1)}^{\otimes N_T}$.
Furthermore, Eq.~(\ref{MEPEdatta}) is a factor $d$ smaller (for $d \gg 1$) when compared to the ${\rm Tr}[\vect{C}]_{\rm MSPE} = d^3/N^2_T$ achieved with the optimal MSPE state of Fig.~\ref{Fig2}(a)~\cite{HumphreysPRL2013}.
The scheme of Fig.~\ref{Fig1}(c) has been also studied when using a QC for the preparation of a ME state, considering multimode Fock~\cite{CiampiniSCIREP2016} as well as Gaussian~\cite{NicholsPRA2018, OhPRR2020} states as input.   

The multimode interferometer of Fig.~\ref{Fig1}(c) has been implemented experimentally in various photonic platforms, in both discrete- and continuous-variable frameworks.
Integrated circuits have been used for the estimation of two~\cite{PolinoOPTICA2019} and three~\cite{ValeriPRR2023} optical phases with pairs of single photon Fock states.
This platform has reached a sensitivity below the classical bound, also implementing various optimal sensing strategies~\cite{CiminiJPJQI2024, CiminiPRApp2021}, see discussion below.
Reference~\cite{GuoNATPHYS2020} has realized the scheme of Fig.~\ref{Fig1}(c) by splitting a displaced squeezed state into $d=4$ modes with a QC.
In this experiment, a strong coherent state provides the phase reference for final homodyne detection, obtained by recombining each mode with the common reference one and measuring the phase quadrature.
This entangled DQS scheme reaches a higher sensitivity for the estimation of the arithmetic average of the phase shifts, $(\theta_1+...+\theta_4)/4$, with respect to a sequential protocol, where the sensing nodes are interrogated with independent squeezed states and using the same total intensity of the squeezed light~\cite{GuoNATPHYS2020}.
Similar to the scheme of Box.~\ref{box:MZI}, a practical advantage is that the ME approach can reach a sub-SQL sensitivity while using a single squeezed state, rather than the $d$ squeezed states necessary in the separable approach. 
The main difference with respect to the Mach-Zehnder sensor networks is the intrinsic requirement of a global measurement scheme.
The experiment~\cite{HongNATCOMM2021} has demonstrated DQS using a four-mode NOON states similar to Eq.~(\ref{psidatta}), with $N=2$ photons, for the estimation of $d=3$ phases.
The multi-mode NOON state is recombined by a QC before final single-photon detection and the corresponding FIM is computed.
The experiment demonstrated that each phase can be estimated with a sensitivity overcoming the SQL.

{\bf Optimizations.}
The estimation bounds discussed previously (in particular the CRB and the MIB) require a large number of measurement repetitions, $m\gg 1$, for their saturation.
However, realistic scenarios are often constrained by resources -- such as time and number of particles~\cite{RubioPRA2020}.
Therefore, optimizing the probe state and measurement setting becomes essential to maximize information acquired from each measurement, also accounting for noise, biases, and the limited operations of the device. 
Devising optimization strategies is particularly demanding in the multiparameter case~\cite{GebhartNRP2023}.

{\it Real-time optimization.}
Adaptive control protocols typically use Bayesian techniques.
The prior knowledge, $P(\vect{\vartheta})$, about the parameter to be estimated is updated after each measurement outcome. 
The Bayes's theorem provides the distribution $P(\vect{\vartheta}\vert \vect{k}) = P(\vect{k}\vert\vect{\vartheta})P(\vect{\vartheta})/P(\vect{k})$, where $P(\vect{k}\vert\vect{\vartheta})$ is the likelihood function and $P(\vect{k})$ provides the normalization.
The posterior $P(\vect{\vartheta}\vert \vect{k})$ quantifies the probability (in the sense of a degree of belief) that $\vect{\vartheta}$ equals the true value of the parameter, $\vect{\theta}$.  
From $P(\vect{\vartheta}\vert \vect{k})$, it is possible to derive the Bayesian covariance matrix, 
\be 
\vect{C}_B = \int d^d\vect{\vartheta} P(\vect{\vartheta} \vert \vect{k}) [\vect{\theta} - \vect{\vartheta}] [\vect{\theta} - \vect{\vartheta}]^T,
\ee
in analogy to Eq. (\ref{Eq.CovMat}), giving the multivariate width of the posterior around $\vect{\theta}$.
While the CRB does not apply in the Bayesian setting, for a sufficiently large number of independent measurements, $m \gg 1$, the posterior becomes a multivariate normal distribution, centered at $\vect{\theta}$ and with $\vect{C}_B = \vect{F}^{-1}/m$~\cite{GebhartNRP2023}. 
The Bayesian approach allows to calculate the uncertainty associated to a specific set of outcomes and to make predictions about future measurement results, guiding the tuning of the probe state, control parameters and measurement observables.
Bayesian protocols, where control phases are progressively adapted to drive the system toward optimal working point, have been implemented in a multiarm interferometer using machine learning~\cite{CiminiPRApp2021, CiminiAP2023} and variational~\cite{ValeriNPJQI2020} adaptive optimization.
Reference~\cite{GebhartPRApp2021} has proposed a Bayesian quantum phase estimation protocol using single qudits in the scheme of Fig.~\ref{Fig1}(c) and a multiple-interrogation protocol with real-time optimization. 
The main obstacles of implementing the Bayesian protocol are the required calibration of the experimental apparatus and the computational resources required for dealing with continuous $d$-dimensional posterior functions. 

{\it Off-line optimization and noise mitigation.}
Designing optimal quantum protocols for specific multiparameter sensing schemes is a complex task that has been only partially addressed in the literature. 
Efforts have been devoted to devise schemes able to overcome the trade-offs due to incompatibility.
A prototypical example is the estimation of parameters $\theta_1$ and $\theta_2$ of the quadrature displacement $e^{i\theta_1 \hat{Q}+ i \theta_2 P}$~\cite{HolevoBOOK1982}, where $P$ and $Q$ are quadrature operators.
In this case, a scheme using symmetric two-mode squeezed state and specific Gaussian measurements~\cite{GenoniPRA2013} can saturate HB~\cite{BradshawPRA2018} and estimate simultaneously the two parameters with high precision, as demonstrated experimentally in an optical system~\cite{SteinlechnerNATPHOT2013}.
A further example is vector field sensing with $N_T$ qubits, which is relevant for magnetometry~\cite{MengNATCOMM2023}.
It consists in estimating the parameters $B$, $\theta$ and $\phi$ that characterize the transformation $e^{i \vect{n}\cdot \vect{J}}$, where $\vect{n} = B (\sin \theta \cos \phi, \sin \theta \sin \phi, \cos \theta)^\top$, $\vect{J} = (J_x, J_y, J_z)$, $J_{h} = \tfrac{1}{2}\sum_{j=1}^{N_T} \sigma_{h}^{(j)}$ and $\sigma_{x,y,z}^{(j)}$ are Pauli operators for the $j$th spin.   
Optimal quantum states that saturate the QCRB have been discussed in Refs.~\cite{BaumgratzPRL2016, HouPRL2020}. 
Vector field sensing has been explored experimentally with photonic qubits~\cite{HouSCIADV2021, XiaNATCOMM2023} showing probe states and measurement schemes~\cite{YuanPRL2016} overcoming incompatibility trade-offs~\cite{ChenNPJQI2024}.

Complex optimization techniques for multiparameter sensing include variational methods~\cite{KaubrueggerPRXQuantum2023, MeyerNPJQI2021}, also implemented experimentally in a multimode interferometer~\cite{CiminiJPJQI2024}, machine learning approaches, which have been used to improve the performance of vector magnetometry~\cite{MengNATCOMM2023}, conic programming for states and measurements~\cite{HayashiNPJQI2024}, and optimal control~\cite{YuanPRA2017}.
Additional goals include optimizing the number of measurement repetitions and extending the range of parameter values that can be estimated with entanglement-enhanced sensitivity.
Finding the optimal trade-off between precision, accuracy, bandwidth and resource consume is important in order to clarify the advantage of entangled multiparameter estimation~\cite{GoreckiPRL2022}.
The challenge further sharpens when including experimental imperfections and decoherence, which are central to designing practical quantum sensors~\cite{SekatskiPRR2020, AlbarelliPRX2022}.
Developing estimation protocols that are robust to noise, such as those incorporating error correction~\cite{ZhuangNJP2020, GreckiQUANTUM2020} and noise-aware estimation algorithms~\cite{GoldbergPRR2023}, is also essential to enhancing precision.
%

{\bf Conclusions.}
The recent vibrant research activity in multiparameter quantum metrology and sensing has significantly advanced our understanding -- both theoretical and experimental -- of a field with foundations dating back to the 1970s \cite{HolevoBOOK1982, HelstromBOOK1976}.
Beyond its potential for groundbreaking advances in precision measurements, the simultaneous estimation of multiple parameters may serve as a bridge between quantum sensing and other quantum technologies, including simulation, cryptography, and computation.
This interdisciplinary cross-fertilization remains largely unexplored, presenting opportunities for future research.
From a theoretical perspective, only relatively simple sensing  configurations have been investigated in details so far. 
It would be important to explore if more complex network geometries and entangled probe states that can provide significant advantages in the presence of imperfections and decoherence.
From the experimental point of view, the challenge is to prove the quantum advantage with respect to current classical protocols in cutting-edge technological applications. \\

{\it Acknowledgments.}
We thank Francesco Albarelli, Marco Barbieri, Animesh Datta, Rafal Demkowicz-Dobrzański, Satoya Imai, Fabio Sciarrino, and Geza T\'oth for comments and suggestions.
This work has been supported by the QuantERA project SQUEIS (Squeezing enhanced inertial sensing), funded by the European Union’s Horizon Europe Program and the Agence Nationale de la Recherche (ANR-22-QUA2-0006);


\end{document}